\documentclass[fleqn,usenatbib]{rasti}

\usepackage{newtxtext,newtxmath}
\usepackage[T1]{fontenc}

\DeclareRobustCommand{\VAN}[3]{#2}
\let\VANthebibliography\thebibliography
\def\thebibliography{\DeclareRobustCommand{\VAN}[3]{##3}\VANthebibliography}

\usepackage{graphicx}	% Including figure files
\usepackage{amsmath}	% Advanced maths commands

\title{Global 21cm Measurement Calibration Methodology}

\author[Bucher et al.]{Martin Bucher,${}^1$ Christian J. Kirkham,${}^{2,3}$ 
Eloy de Lera Acedo,${}^{2,3}$ Dirk I. L. de Villiers${}^4$ and Saurabh Pegwal${}^4$ 
\\
$^{1}$Laboratoire APC, Université Paris Cité/CNRS, F-75013 Paris, France\\
$^{2}$Astrophysics Group, Cavendish Laboratory, J.J. Thomson Avenue, Cambridge, CB3 0HE, UK\\
$^{3}$Kavli Institute for Cosmology, Madingley Road, Cambridge, CB3 0HA, UK\\
${}^4$Department of Electrical and Electronic Engineering, Stellenbosch University, Stellenbosch, South Africa%\\
}

\date{Accepted XXX. Received YYY; in original form \today }

\pubyear{2015}

\begin{document}
\label{firstpage}
\pagerange{\pageref{firstpage}--7}
\maketitle

\begin{abstract}
21cm global signal observations present a unique set of calibration challenges owing to the absolute character 
of the required measurement. Since differential measurements on the sky cannot be used for observing the global signal,
typically a number of calibration sources with differing noise temperatures and source impedances are used to
determine the four noise parameters and the power gain of the amplification chain. Because of the broadband nature of the measurement,
the antenna impedance varies with frequency in a manner different from the calibration sources, so that the simplest
three-way Dicke switching strategy is not adequate.
We present a self-contained and explicit derivation of the calibration equations and reconcile expressions
from the early global 21cm observation literature with the results obtained following the 
amplifier noise representation formalism commonly used in the electrical engineering literature. We also present a condition in terms of M\"obius geometry on the complex $\Gamma $- (or $Z$-) plane 
defining the choice of calibration source impedances required. 
\end{abstract}


%%%%%%%%%%%%%%%%%%%%%%%%%%%%%%%%%%%%%%%%%%%%%%%%%%

%%%%%%%%%%%%%%%%% BODY OF PAPER %%%%%%%%%%%%%%%%%%

\section{Introduction}

The distortions of the extreme low-frequency tail 
of the CMB blackbody spectrum arising from the absorption or emission
of the redshifted 21cm HI hyperfine transition line offer a new
probe of our universe at the intermediate redshifts between recombination
and today. This distortion as a function of frequency averaged over
the celestial sphere, known as the global 21cm signal,  
depends on both the HI density and the spin temperature. Its measurement
will help constrain the history of the creation of the first stars and
quasars as well as other processes during this epoch. 
[See for example \citep{global21cmTheoryReview,cohen2018charting} and 
references therein for a discussion.]
When Galactic 
foregrounds are neglected, the global 21cm signal corresponds to
the $Y_{00}$ spherical harmonic basis function whereas the differential signal corresponds
to the coefficients of the higher ($l\ge 1$) multipoles. The expected global 
signal is much larger than the differential signal. This advantage,
however, is offset by the need for an absolute calibration. 

The EDGES collaboration \citep{edgesDetectionPaper}
has claimed a statistically significant detection of a nonzero
signal. However most theoretical models, while predicting
a signal centered near the same frequency, predict
an absorption feature having 
an amplitude at least twice as small.
Other experiments [e.g., REACH  \citep{reachNaturePaper}, SARAS
\citep{saras3,saras2}, PRIZM \cite{philip}, and MIST \cite{Monsalve:2023lvo} ] 
are underway to search for a global signal
in the same frequency range. Some have reported preliminary
results [SARAS 2 \citep{saras2-bis}]. 

The measured brightness temperature includes
the above extragalactic signal as well as a much larger foreground 
contribution from synchrotron emission 
of our Galaxy and other smooth contributions (e.g., extragalactic synchrotron, free-free emission). 
This Galactic foreground must be removed 
or otherwise accounted for as part of any
analysis aimed at extragalactic science extraction. 
In practice the measurement consists of many 
measurements taken over parts of the whole sky, which are combined. 
These measurements are taken using an antenna with 
little directivity (owing to the large wavelengths involved). But for 
purposes of this Note, we ignore 
all complications involving the character of the antenna pattern on the sky.
Here we focus on the signal emanating from the antenna and 
entering the receiver, ignoring for the moment the important issues
of sky modelling, modelling of the antenna directivity, 
corrections from ground pick-up and
from RFI near the horizon, and more general data analysis issues. 
Instead the point of departure here is an equivalent 
circuit source model for the antenna 
output (including possible Ohmic losses internal 
to the antenna). We focus on the 
measurement of the absolute frequency spectrum 
of this incident signal, which may be expressed
as a frequency dependent Rayleigh-Jeans brightness temperature $T(\nu ).$

\begin{figure}
 \begin{center}
        \includegraphics[width=0.9\columnwidth,  trim = 50 10 50 0, clip]%
%{receiverSchematic.png}\\
{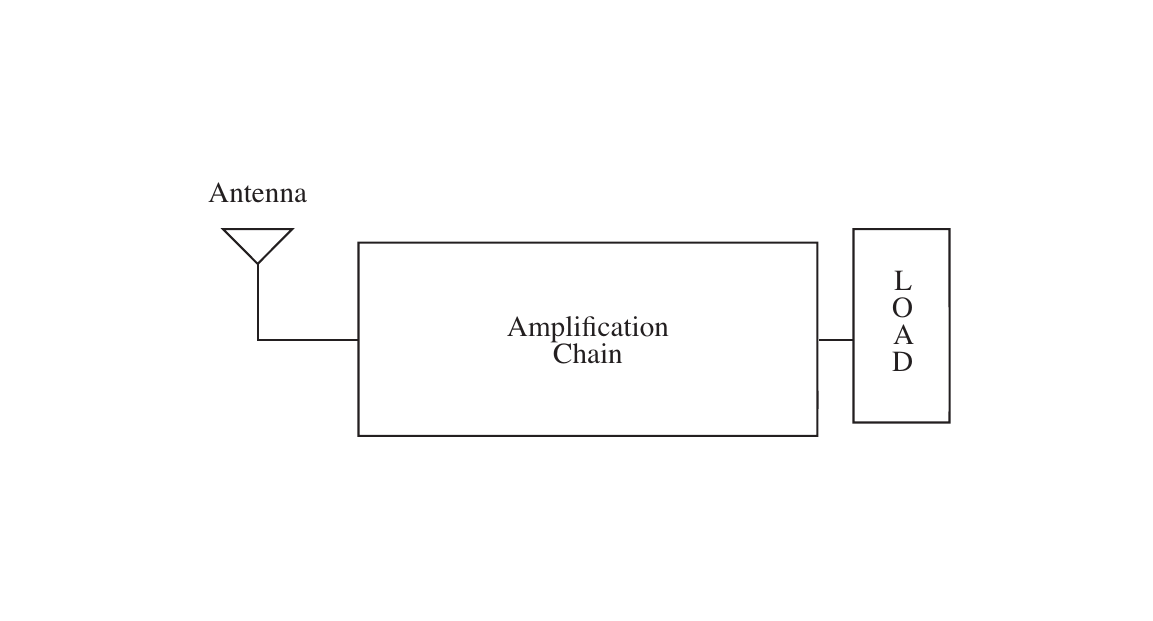}\\[-25pt]
(a)\\[-10pt]
        \includegraphics[width=0.99\columnwidth,  trim = 50 10  50 0, clip]%
%{singleSourceEquivCircuit.png}\\
{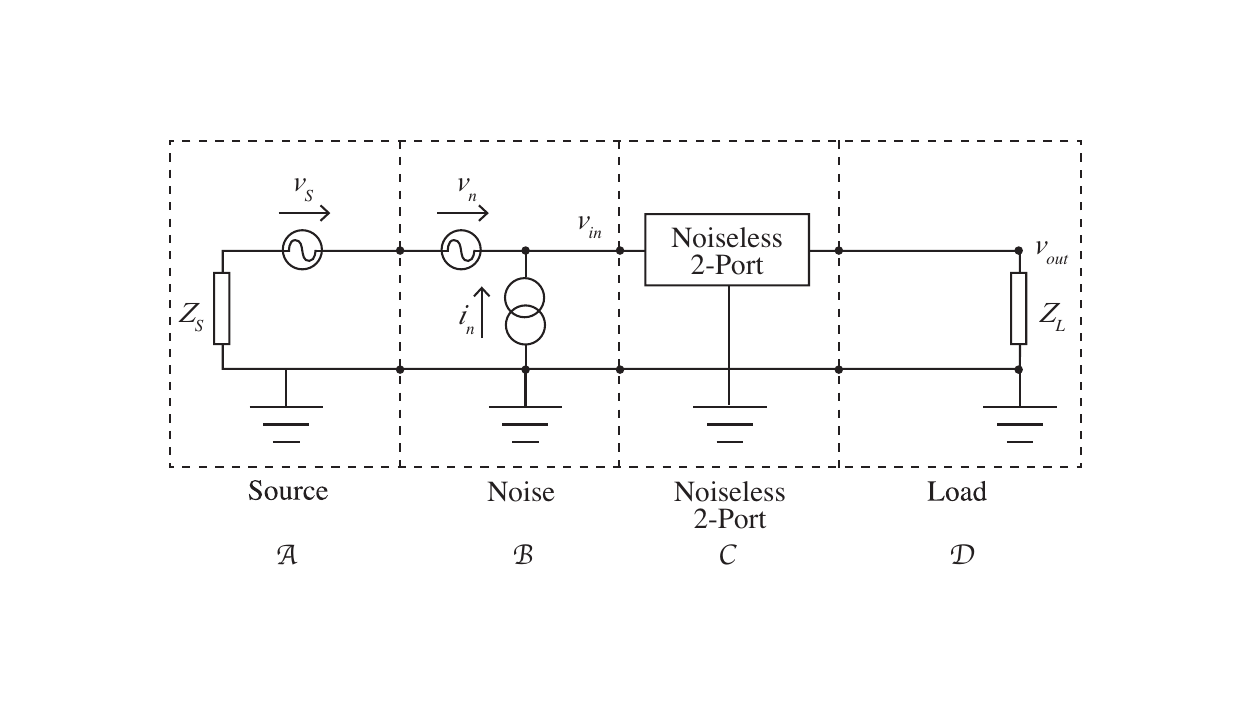}\\[-20pt]
(b)\\
    \end{center}
    \caption{{\bf Receiver Schematic (a) and its Equivalent Circuit (b).}
The signal enters through the antenna, represented by
a Thévenin equivalent circuit ($\mathcal{A}$). The noisy amplifier
$(\mathcal{B} + \mathcal{C})$
is a noisy linear 2-port device, which may be represented
as noiseless linear 2-port device 
$\mathcal{C}$
with a pair of noise sources 
$\mathcal{B}$
placed in front of its input   
[Haus et al. (1960)  
and references therein], which drives a load
$\mathcal{D}$ of impedance $Z_L.$ The voltage $v_\textrm{out}$ is digitized
by an Analogue-to-Digital Converter (ADC). If $Z_S,$ the scattering and noise parameters of the noisy
amplifier, and the load impedance $Z_L$ were stable in time and known
with adequate precision, an absolute measurement could be made with a 
setup of this sort. In our analysis, we shall find it convenient
to replace $(\mathcal{A} + \mathcal{B})$ with a Thévenin equivalent
source, even though separation between  
$\mathcal{B}$ and $\mathcal{C}$ in the amplifier is an artifact of
the representation. Physically, noise generation is distributed throughout
the amplifier. However, network circuit analysis regards components
as black boxes, and one is not allowed to look inside.  
}
    \label{fig:newRecSchem}
\end{figure}

As shown schematically in Fig.~\ref{fig:newRecSchem}(a), 
the sky signal emanating from the antenna output passes through an amplification chain whose details
need not concern us here, after which it is digitized by an ADC.
In the linear approximation (so that different frequencies can be analysed separately), 
the analogue part of the setup can be modelled as the equivalent circuit shown in 
Fig.~\ref{fig:newRecSchem}(b).\footnote{AC/microwave network circuit theory is 
expounded in texts such as \cite{pozar} where the discussion of amplifier noise 
focuses on the special case where impedances are matched. \cite{collin} provides
a more in depth discussion. The general case
was worked out in the late 1950s and 1960s in the references cited below.
In network circuit theory, there are no reference planes: an $N$-port device is a black box, and the various
representations of the $N$-port express the $N$ linear constraints on the $2N$ variables at the ports
(either $N$ voltages and $N$ currents, or alternatively $N$
incoming and $N$ outgoing travelling wave amplitudes with respect to some characteristic impedance).
Connections between ports are idealized to have vanishing length, and finite lengths of transmission line
are represented as 2-port devices. 
}
The amplifier unit is a noisy two-port device, here driving a load having an 
impedance $Z_L.$ The output voltage $v_\textrm{out}(t)$ (after filtering) is fed into an ADC, so that
the rest of the processing and analysis is carried out digitally. 
If we knew the antenna impedance $Z_a,$
the amplifier input impedance $Z_\textrm{in},$ 
and the amplifier gain and noise properties 
with sufficient precision, and moreover if these were stable
in time, there would be no need to calibrate 
and the setup in 
Fig.~\ref{fig:newRecSchem}(a) would suffice.  
In practice, however, these quantities are not known 
with sufficient precision, and moreover they vary in time.
Consequently, we must calibrate, and depending on the time 
stability of the setup, calibration must occur periodically. 

The study of the theory of noisy linear two-ports and their generalizations dates
back to the 1950s with the work of \cite{rothe}
[see also \cite{bauer} (in German)], \cite{hausAdler}, 
\cite{reprNoisyTwoPorts}. (See also \cite{bosma}.)
Although some of this work includes more general results (for representing a noisy $N$-port 
where $N>2),$ for our purposes the central result proven is that a noisy linear amplifier
can be represented by a noiseless 2-port with a voltage source-current source pair 
placed in front of the amplifier input.\footnote{For a nonunilateral amplifier (a unilateral
amplifier is a singular but important case), the noise sources could equally well be placed
following the output port), but for most applications this representation is less meaningful.}

\begin{figure}
 \begin{center}
        \includegraphics[width=0.9\columnwidth,trim = 100 30  100 30, clip]%
   %{2p_noise_source.png}
   {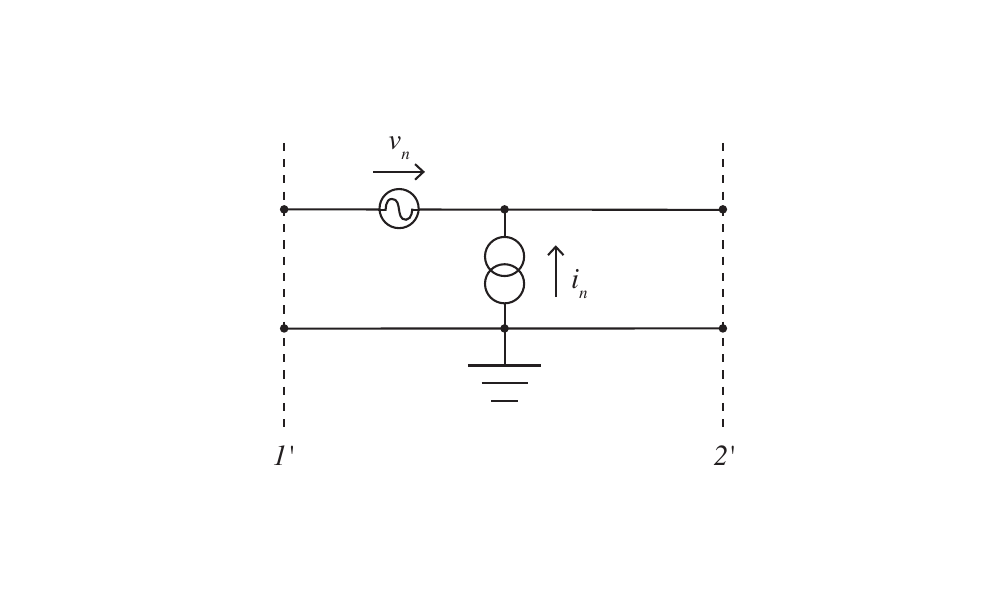}\\[-25pt]
   (a)\\[-20pt]
        \includegraphics[width=0.9\columnwidth,trim = 100 30  100 30, clip]%
   %{2p_noise_source.png}
   {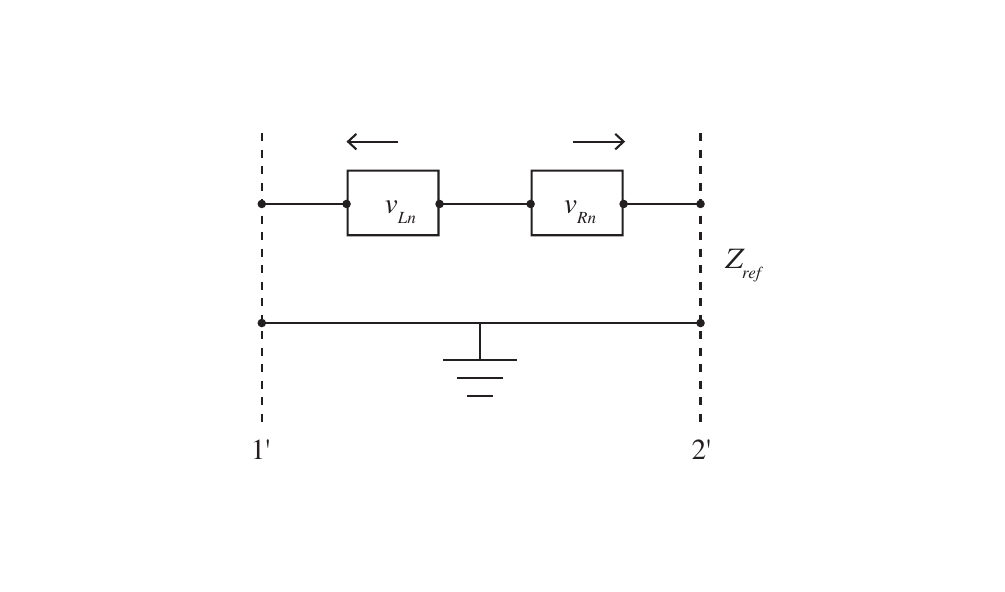}\\[-25pt]
   (b)
    \end{center}
    \caption{{\bf Two-Port Noise Element.}
       The above blocks represent two equivalent ways to represent noise added to a port.
       Here the indicated variables are complex Gaussian variables, which in the general case
       are correlated, but they could also be fixed complex numbers, in which case much of the 
       math would be the same. 
       In (a) we see a voltage- and current-source pair, whereas in (b) we see 
       a travelling wave representation in terms of left- and right-moving waves with
       respect to a real reference characteristic impedance 
       $Z_\textrm{ref},$ which is arbitrary. 
       In both cases, the variables indicate jump conditions between the two inputs.
       The homogeneous part of the linear description of this 2-port is 
       simply $\mathbf{S}=\mathbf{I}$ corresponding to complete transparency 
       of the sources.}
    \label{fig:2pNoiseSource}
\end{figure}

It is also possible to use a travelling wave, or power wave, formalism, described in detail
by \cite{kurokawa}, so that as indicated in 
Fig.~\ref{fig:newRecSchem}
pre-placed voltage and current sources are replaced by pre-placed left- and right-moving
travelling wave noise sources. 
This completely equivalent formulation was worked out by
\cite{penfield}
and later by \cite{meysPaper}  %{\color{blue}[180 citations]}]. 
The two formulations, either in terms of voltage and current sources
shown in Fig.~\ref{fig:2pNoiseSource}(a)
or in terms of travelling wave sources 
shown in Fig.~\ref{fig:2pNoiseSource}(b),
are mathematically equivalent.  
The travelling wave representation is 
always with respect to some real arbitrary reference characteristic impedance $Z_\textrm{ref},$
which is a purely mathematical construct and does not necessarily 
correspond to any physical transmission line in the circuit. 
A test of whether an expression for a physically observable quantity
is consistent is whether this quantity is invariant under changes of the 
$Z_\textrm{ref}$
used to define the travelling power waves.

As already stated, if all the noise, amplifier, and load parameters were perfectly known, the setup
in Fig.~\ref{fig:newRecSchem}(a) would suffice, but because these parameters are not known 
with sufficient accuracy and also drift with time, periodic recalibration is needed. This is
known as Dicke switching \citep{dickePaper}, which is a term encompassing 
a wide variety of setups.  

\begin{figure}
 \begin{center}
        \includegraphics[width=0.9\columnwidth,  trim = 0 10 0 70, clip]%
{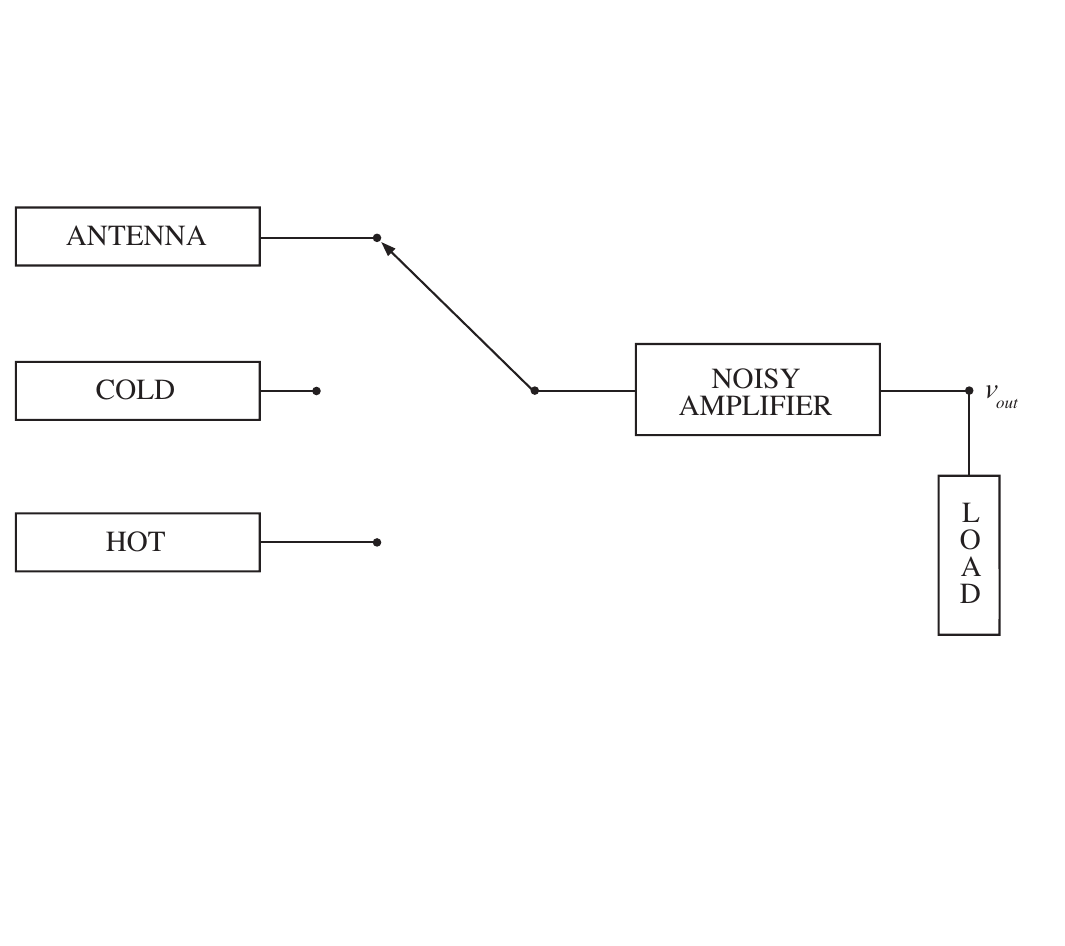}\\[-50pt]
(a)\\[-25pt]
        \includegraphics[width=0.9\columnwidth,  trim = 10 10 10 0, clip]%
{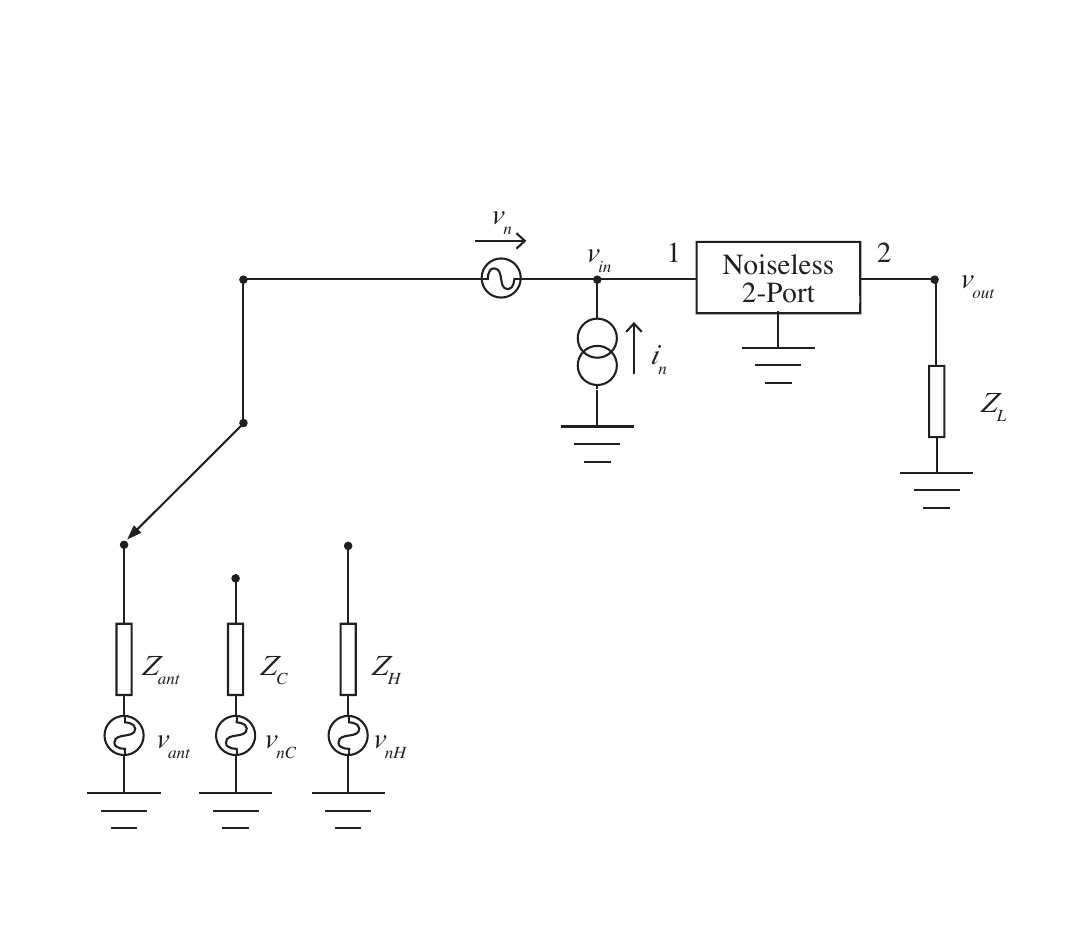}\\[-10pt]
(b)
    \end{center}
    \caption{{\bf Three-Way Dicke Switching.}
For the special case where the antenna source impedance 
exactly matches the source impedances of the hot and cold
calibration sources, it is possible to calibrate even when the 
properties of the amplifier and load are unknown.  This setup,
shown schematically in (a) and in terms of an equivalent circuit in (b), 
is analogous to commercially available 
noise figure/gain measurement devices using the $y$-intercept method
to determine
$G$ and $T_\textrm{amp,\,n},$ so
that the source noise temperature can be measured when 
the device source impedance is exactly matched to the calibration
source impedance. However for a source of an arbitrary source 
impedance, this setup is insufficient, as knowledge
of all four noise parameters of the amplifier is required.
The faster amplifier characteristics vary with time, 
the faster one must switch between the test source and
the calibration sources.
}
    \label{fig:threeWayDicke}
\end{figure}

\begin{figure}
 \begin{center}
        \includegraphics[width=0.9\columnwidth,  trim = 60 10 0 20, clip]%
%{sixWayDS-EC.png}
{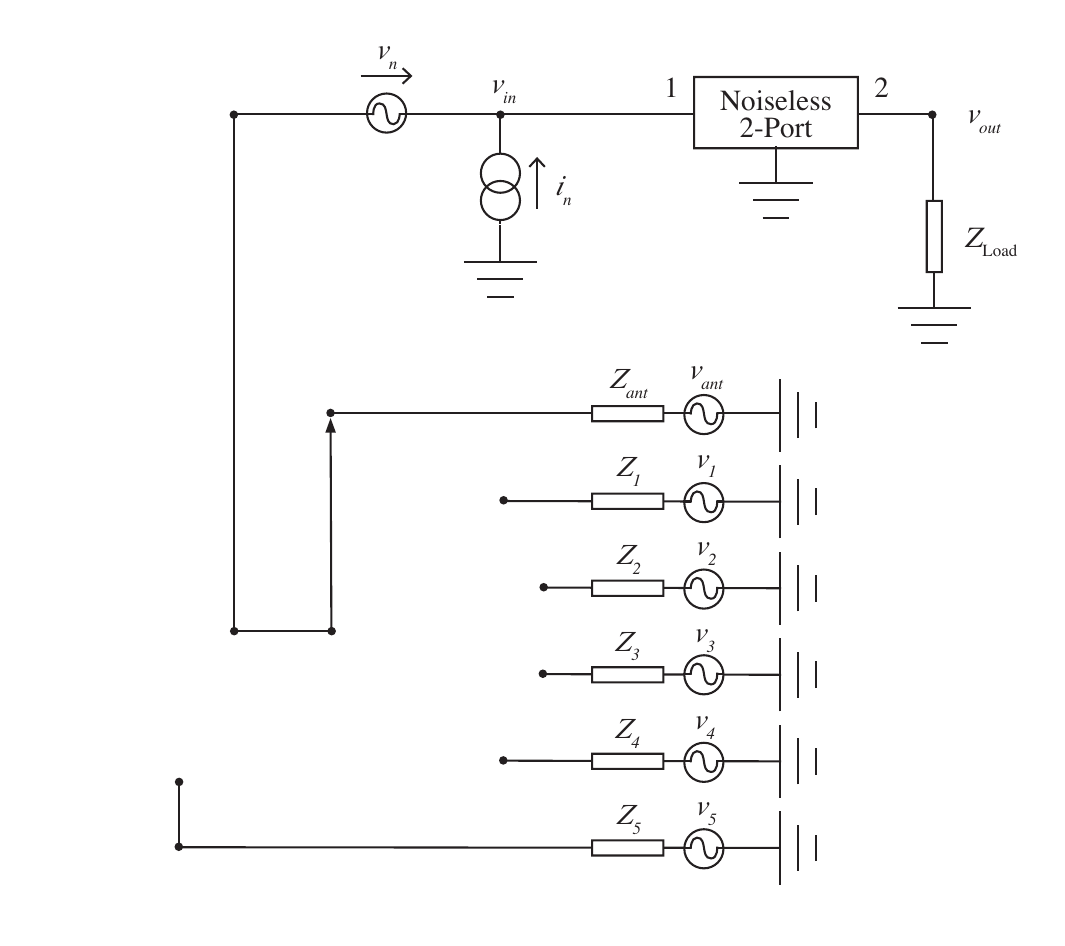}
    \end{center}
    \caption{{\bf Six-Way Dicke Swtiching.}
The shortcomings of three-way Dicke switching
may be remedied by using at least five independent calibration
sources as shown, so that all four noise parameters and the gain
of the amplifier (for the particular load impedance) can be 
determined, and thus the noise temperature of  
a test source of arbitrary source impedance
can be measured.
While at least five calibration sources are needed, frequent switching between a subset 
of them and the test source may suffice, for example if only the gain suffers significant
time variation.
}
    \label{fig:sixWay}
\end{figure}

Figure \ref{fig:threeWayDicke}(a) illustrates an oversimplified 
calibration scheme, which in addition to the test source (here the antenna)  
includes two calibration sources, a cold Johnson noise source (at temperature 
$T_C$) and a hot source  (at $T_H$), which we must assume to have
the same source impedance as the test source at all frequencies. 
Such impedance matching is not feasible, especially for a broadband 
antenna as is required for global 21cm observations, so we will have to
replace this Dicke switching scheme with a more sophisticated 6-way 
Dicke switching scheme, as explained further below
(see Fig.~\ref{fig:sixWay}).
The 3-way scheme is inadequate because
the noise added by the amplifier is source impedance dependent. 
Figure \ref{fig:threeWayDicke}(b) provides an equivalent circuit 
of this scheme. 

Following a naive analysis of the three-way Dicke switching, we 
write 
\begin{equation}
\begin{aligned}
P_\textrm{out}
&=c\left[ T_\textrm{amp,\,n}+T_\textrm{source} \right] 
\end{aligned}
\label{eqn:first}
\end{equation} 
where $T_\textrm{amp,\,n}$ is the noise generated within the amplifier, $T_\textrm{source}$
is one of the three sources, and $c$ is a constant of proportionality.
Here the noise generated within the amplifier is
an additional additive input, where we 
tacitly assume that the amount of noise added does not depend on
the nature of  what is placed before the amplifier input.
Here we do not care exactly how $P_\textrm{out}$ is defined;
we rely solely on the linearity expressed 
in eqn.~(\ref{eqn:first})
and find that
\begin{equation}
T_\textrm{source}=T_C+
\left(
\frac{T_H-T_C}{P_H-P_C}
\right) 
(P_\textrm{source}-P_C).
\end{equation}

We take a closer look at the three-way Dicke switching, using 
the equivalent circuit in Fig.~\ref{fig:threeWayDicke}(b) 
and relaxing the assumption that the three source impedances match
at all frequencies. 
In experiments such as REACH, EDGES, and SARAS, because of the broadband 
design of the antenna, the antenna impedance 
as a function of frequency has a complicated structure and matching
with the calibration sources is not feasible. 

We may replace each of the three sources with a single voltage
source by constructing a Th\'evenin equivalent circuit for everything
in front of the amplifier input to obtain\footnote{%
This would be the voltage at the amplifier input when 
$Z_\textrm{in}=\infty ,$ but for finite $Z_\textrm{in}$ there is an additional $Z_\textrm{in}/(Z_S+Z_\textrm{in})$ factor.}
\begin{equation}
v_\textrm{in}=v_\textrm{S}+v_{na}+Z_{S}~i_{na},
\end{equation}
so that 
\begin{equation}
\left\langle \vert v_\textrm{in}\vert ^2\right\rangle
=
\left\langle \vert
v_\textrm{S}
\vert ^2\right\rangle
+
\left\langle \vert 
v_{na}+Z_{S}i_{na}
\vert ^2\right\rangle 
\end{equation}
where we have assumed a lack of correlation between 
the input source fluctuations and 
the amplifier noise fluctuations.
Here $S$ stands for source, which can be either the antenna 
or one of the calibration sources, and 
$na$ denotes the noise generated by the amplifier. Expanding the second term,
we obtain 
\begin{equation}
\left\langle 
v_{na}^*
v_{na}^{\phantom{*}} 
\right\rangle 
+Z_{S}^{\phantom{*}} \left\langle 
v_{na}^* 
i_{na}^{\phantom{*}} 
\right\rangle
+Z_{S}^* \left\langle 
i_{na}^* 
v_{na}^{\phantom{*}} 
\right\rangle
+ Z_{S}^{\phantom{*}} Z_{S}^* \left\langle 
i_{na}^* 
i_{na}^{\phantom{*}} 
\right\rangle . 
\label{eqn:ampNoise}
\end{equation}
The noise parameters are the four expectation values, 
two of which are real and the other two of which are 
complex conjugates of each other. In terms of the signal
at the output, we have
\begin{equation}
\begin{aligned}
&
\left\langle \vert
v_\textrm{out}
\vert ^2 \right\rangle
=
\left\vert G_V\right\vert ^2 
\left\vert 
\frac{Z_\textrm{in}}
{
Z_\textrm{in}+Z_S
}
\right\vert ^2 \times 
\biggl[
\left\langle \vert
v_\textrm{S}
\vert ^2 \right\rangle
\\
&\quad +\Bigl( 
\left\langle v_{na}^* v_{na}^{\phantom{*}} \right\rangle 
+Z_{S} 
\left\langle v_{na}^* i_{na}^{\phantom{*}} \right\rangle 
\\
&\qquad 
+Z_{S}^* 
\left\langle i_{na}^* v_{na}^{\phantom{*}} \right\rangle
+ Z_{S}^{\phantom{*}} Z_{S}^* 
\left\langle i_{na}^* i_{na}^{\phantom{*}} \right\rangle 
\Bigr) 
\biggr] . 
\end{aligned}
\end{equation}
Here $G_V$ and $Z_\textrm{in}$ are not intrinsic properties 
of the amplifier 2-port\footnote{The linear properties of a noiseless amplifier (or more
generally any 2-port) is represented by 8 independent real parameters
in the form of a $2\times 2$ complex matrix, the different representations
including the $S$ matrix, the $T$ matrix (i.e., the $ABCD$ matrix), 
the $Z,$ and the $Y$ matrix. See 
\cite{pozar} for the transformations between these representations.} but 
also depend on the impedance $Z_L$ of the load
connected to the amplifier output. 
For a fixed load impedance $Z_L$ (or $\Gamma _L$ when expressed as a reflection coefficient), the receiver
input reflection coefficient is given by
\begin{equation}
\Gamma _\textrm{in}=
S_{11}
+
\frac{
S_{12}
\Gamma _L
S_{21}
}{
1-S_{22}\Gamma _L
}
\label{eqn:gammaIn}
\end{equation}
as can be derived by summing over reflections.
The voltage gain is given by
\begin{equation}
G_V=
\frac{
1+\Gamma _L
}{
1+\Gamma _\textrm{in}
}
\frac{S_{21}
}{
1-\Gamma _LS_{22}
},
\end{equation}
and the power gain\footnote{A variety
of definitions 
serving different purposes
exist for the amplifier power gain. 
The expression here is the so-called `actual' or `operating'
power gain for the load impedance $Z=Z_L.$
This gain depends on $\Gamma _L$ but not on $\Gamma _S.$
This definition does not reflect the diminution
of the power delivered to the amplifier input compared to the power available 
from the source due to impedance mismatch.}
is given by
\begin{equation}
G_\textrm{P}
=\frac{
P_\textrm{out}
}{
P_\textrm{in}
}
=
\frac{
\left\vert
S_{21}
\right\vert ^2
\left(1- 
\left\vert \Gamma _L\right\vert ^2\right)}{
\left\vert 1- S_{22}\Gamma _L \right\vert ^2
\left(
1- \left\vert \Gamma _\textrm{in}\right\vert ^2
\right)
}.
\end{equation}

For the calibration sources
in thermal equilibrium (i.e., with no DC bias currents), Johnson noise is the only contribution to
the source noise, and 
$\left\langle \vert v_\textrm{S} \vert ^2 \right\rangle =4k_BT~\textrm{Re}(Z_S)\, (\Delta B).$
Calibration involves solving for the five real parameters:
the amplifier gain 
$\left\vert G_V\right\vert ^2,$ and the four noise parameters,
$\left\langle v_{na}^* v_{na}^{\phantom{*}} \right\rangle ,$ 
$\left\langle v_{na}^* i_{na}^{\phantom{*}} \right\rangle ,$ 
$\left\langle i_{na}^* v_{na}^{\phantom{*}} \right\rangle ,$
and 
$\left\langle i_{na}^* i_{na}^{\phantom{*}} \right\rangle .$ 
This is why we need five independent calibration sources.
We require five equations to solve for five unknowns. Moreover
these five equations must be independent, which requires
that the determinant of the matrix 
\begin{equation}
\begin{pmatrix}
1& Z_1 & {Z_1}^* & {Z_1}^*Z_1 \\
1& Z_2 & {Z_2}^* & {Z_2}^*Z_2 \\
1& Z_3 & {Z_3}^* & {Z_3}^*Z_3 \\
1& Z_4 & {Z_4}^* & {Z_4}^*Z_4
\end{pmatrix}
\end{equation}
does not vanish. This last condition has a geometric 
interpretation in terms of Möbius 
geometry,\footnote{Möbius geometry deals with properties of points on
the Riemann sphere invariant under 
Möbius transformations (i.e., global conformal transformations).
In Möbius geometry any three distinct points define a generalized 
circle (either a circle or an infinite line, which may be regarded as a degenerate
case of a circle). [See for example \cite{ahlfors} for a discussion.]}
which can be expressed either on the $Z$-plane or on 
the $\Gamma $-plane (i.e., Smith chart). The determinant above 
vanishes if and only if the four points lie on a common circle. On the complex
plane an equation of the form $a+bZ+cZ^*+dZ^*Z=0$ defines a circle,
and if the homogeneous system of equations defined by this matrix has a nonzero
solution, $Z_1,$ $Z_2,$ $Z_3,$ and $Z_4,$ lie on a common circle.\footnote{Any 
row in the matrix may be re-expressed in terms of the admittance $Y,$
by  making the row replacement $(1, Z, Z^*Z)\mapsto (Y^*Y, Y^*, Y, 1).$ This transformation
allows for the representation of the otherwise singular open circuit.
When $Z$ (or $Y$) are understood projectively as lying
on a Riemann sphere, singularities do not arise
\citep{harris}.}
The property clarifies that the calibration source 
impedances cannot all lie on the unit circle on the Smith chart: 
at least one Z must have a positive real part. Likewise, 
the  $Z$'s cannot all lie on the real axis of the Smith chart:
at least one $Z$ must have a nonzero reactance. This is an 
analysis of the linear part of the system of five equations, but
we must also determine $\left\vert G_V\right\vert ^2$ 
and this requires at least two impedances with a real part
at different temperatures.\footnote{One may also use avalanche diode
noise sources, which are characterized by an equivalent Johnson noise
temperature.} 
If we have two identical impedances with a positive real part
at different temperatures, we may determine 
$\left\vert G_V\right\vert ^2$ by solving a linear equation,
and once this quantity is known, the noise parameters can be 
determined by solving a linear equation. But of course
the general case, where one must solve for five parameters
determined by a nonlinear equation, is numerically trivial. 
The six-way Dicke switching setup sketched in Fig.~\ref{fig:sixWay} 
remedies the shortcomings 
of the three-way Dicke switching (see Fig.~\ref{fig:threeWayDicke}) provided that the impedances
chosen are nondegenerate as described above.

Following \cite{penfield} and \cite{meysPaper}, we 
now sketch how the above calculation would be 
carried out in terms of the Kurokawa travelling wave, 
or power wave, 
formalism, where the voltage and current variables
are replaced with the travelling wave variables \citep{kurokawa}
\begin{equation} 
\begin{aligned} 
a_R=\frac{(v+Z_\textrm{ref}\,i)}{2\sqrt{Z_\textrm{ref}}},\qquad  
a_L=\frac{(v-Z_\textrm{ref}\,i)}{2\sqrt{Z_\textrm{ref}}},
\end{aligned} 
\label{eqn:powerWaveDef}
\end{equation} 
where the normalization is such that 
$\langle \vert a_L\vert ^2 \rangle (\Delta B)$
and 
$\langle \vert a_R\vert ^2 \rangle (\Delta B)$ correspond to
the power propagating leftward and rightward 
when interference effects are ignored. 
For our setup, the right-moving power wave amplitude is given 
by\footnote{Here 
we have defined $A_S$ so that $\left\langle \vert A_S\vert ^2\right\rangle $ 
corresponds to the available power per unit frequency when the load impedance is
perfectly matched for maximal power transfer, hence 
the presence of the 
$\sqrt{1-\vert \Gamma _S\vert ^2}$ 
factor.} 
\begin{equation}
a_R=
\frac{
A_R+\Gamma _SA_L+\sqrt{1-\vert \Gamma _S\vert ^2}A_S
}{
1-
\Gamma _\textrm{in}\Gamma _S
}
\end{equation}
where $A_L$ and $A_R$ are the amplitudes of the left-moving
and right-moving power wave noise sources (i.e., jump conditions) 
and $A_S$ corresponds to the power wave source amplitude. It follows that
\begin{equation}
\begin{aligned}
P_\textrm{in}&=
\left( 1-\vert \Gamma _\textrm{in}\vert ^2\right)
\left\langle
\vert a_R\vert ^2
\right\rangle \\
&=
\frac{
\left( 1-\vert \Gamma _\textrm{in}\vert ^2\right)
}{
\left\vert
1-\Gamma _\textrm{in}\Gamma _{S}
\right\vert ^2} \biggl[
\left( 1-\vert \Gamma _S\vert ^2\right)
\left\langle \vert A_S\vert ^2 \right\rangle
\\
&\quad +
\left\langle \vert A_R\vert ^2 \right\rangle
+
\vert \Gamma _S\vert ^2
\left\langle \vert A_L\vert ^2 \right\rangle
\\
&\quad +
\Gamma _S \left\langle A_R^*A_L^{\phantom{*}} \right\rangle
+\Gamma _S^* \left\langle A_L^*A_R^{\phantom{*}}\right\rangle
\biggr] ,
\end{aligned}
\label{eqnPinc}
\end{equation}
which may be re-expressed as the brightness temperature corresponding to 
the incident power spectral density absorbed by the amplifier
\begin{equation}
\begin{aligned}
T_\textrm{in}
&=
\frac{
\left( 1-\vert \Gamma _\textrm{in}\vert ^2\right)
}{
\left\vert
1-\Gamma _\textrm{in}\Gamma _{S}
\right\vert ^2} \biggl[
\left( 1-\vert \Gamma _S\vert ^2\right) T_S
\\
&\quad 
+T_R
+\vert \Gamma _S\vert ^2 T_L 
+2\vert \Gamma _S\vert \Bigl( T_\textrm{cos}\cos \phi +T_\textrm{sin}\sin \phi \Bigr) \biggr] 
\end{aligned}
\label{eqnTinc}
\end{equation}
where 
\smash{
$T_\textrm{cos}=\textrm{Re} \left( \left\langle A^*_RA_L^{\phantom{*}} \right\rangle \right) ,$}
\smash{
$T_\textrm{sin}=-\textrm{Im}\left( \left\langle A^*_RA_L^{\phantom{*}} \right\rangle \right) ,$}
and 
\smash{$\Gamma _S=\vert \Gamma _S\vert (\cos \phi +j\sin \phi ) .$}
The frequency dependent temperatures 
$T_L$
and 
$T_R$
have a physical interpretation; they are, respectively, the power spectral densities of the
left- and right-moving Kurokawa virtual noise wave sources placed before the
virtual noiseless amplifier input, and 
$z=
(T_\textrm{cos}+jT_\textrm{sin})/
\sqrt{T_LT_R}
$
where $\vert z\vert \le 1$ is their dimensionless complex correlation coefficient. 

In coefficient of $T_S$ in the expression for $T_\textrm{in}$
in eqn.~(\ref{eqnTinc}), we recognize the familiar impedance mismatch
factor 
\begin{equation}
\begin{aligned}
M&=\frac{
\left( Z_S+Z_S^* \right)
\left( Z_L+Z_L^* \right)
}{
\left\vert
Z_S+Z_L
\right\vert ^2
}\\
&=\frac{
\left( 1-\left\vert \Gamma _S\right\vert ^2\right)
\left( 1-\left\vert \Gamma _L\right\vert ^2\right)
}{
\left\vert 1- \Gamma _S\Gamma _L\right\vert ^2
},
\end{aligned}
\end{equation}
the expression on the second line being invariant
under changes in $Z_\textrm{ref}$ as it must be.

The calibration procedure described above assumes that all the impedances 
and scattering parameters are known.
Care must be taken to include the contributions from cables and switches 
and to ensure that a common reference point is used. In the REACH experiment
a VNA is built into the front end to measure all the scattering parameters
periodically as part of the calibration procedure \citep{roque}.

\section{Discussion} 

We compare the results derived above with 
certain %conflicting 
results that have appeared 
in the 21 cm global experiment
literature, highlighting
the discrepancies and explaining the reasoning that
has led to these differences.
When the methods first developed in the 
earlier electrical engineering literature are
applied, the results above in eqns.~(\ref{eqnPinc}) and (\ref{eqnTinc})
follow.

Eqn. (8) in \cite{rogersA} 
[hereafter R\&B], which reads as follows
\begin{equation}
\begin{aligned}
T_\textrm{rec}&=
 T_\textrm{sky}(1-\vert \Gamma _{a}\vert ^2 )~\vert F\vert ^2\\
&+T_u~\vert \Gamma _{a}\vert ^2 ~ \vert F\vert ^2\\
&+\Bigl( T_c\cos \phi + T_s\sin \phi \Bigr)~ \vert \,\Gamma _a\,\vert ~\Big[ \,\vert F\vert \,\Bigr] \\ 
&+\Bigl[ \,T_0\, \Bigr] ,
\end{aligned}
\label{rbEqn8}
\end{equation}
conflicts with the analysis in this paper,  
which instead finds that
\begin{equation}
\begin{aligned}
T_\textrm{in}&=
 T_{S}(1-\vert \, \Gamma _{S}\,\vert ^2 )~\vert F\vert ^2\\
&+T_L~\vert \Gamma _{S}\vert ^2 ~ \vert F\vert ^2\\
&+2\Bigl( T_\textrm{cos}\cos \phi + T_\textrm{sin}\sin \phi \Bigr) ~
\vert \Gamma _S\vert ~
\Bigl[ \,\vert F\vert ^{2}\,\Bigr] \\
&+ T_R  \Bigl[ \, \vert F\vert ^2\,\Bigr] . 
\end{aligned}
\label{RBcorrected}
\end{equation}
The differences have been highlighted using square brackets indicating 
the additions to eqn.~(\ref{rbEqn8}) required 
to obtain agreement with 
eqn.~(\ref{eqnTinc}).\footnote{Eqn.~(\ref{rbEqn8}) copies verbatim the subscript labels of
R\&B, whereas in eqn.~(\ref{RBcorrected}) the notation
has been rendered consistent with the previous section, for example by reflecting that the 
source is not necessarily an antenna but can also be a calibration source. Thus we 
replace $T_\textrm{sky}$ with $T_S,$ and $\Gamma _a$ with $\Gamma _S.$
Furthermore, we replace $T_u$ and $T_0$ with the more suggestive $T_L$ and $T_R,$
respectively, and we also replace $\Gamma _l$ with $\Gamma _\textrm{in},$ which is the 
amplifier input reflection coefficient.}
The expressions
$T_\textrm{rec}$ and $T_\textrm{in}$ 
in eqns.~(\ref{rbEqn8}) and (\ref{RBcorrected}), respectively, 
represent the same physical quantity: the power spectral density absorbed 
at the amplifier input expressed as a temperature. 
Here $\vert F\vert ^2=(1-\left\vert \Gamma _\textrm{in}\right\vert ^2)/\left\vert 1-\Gamma _S\Gamma _\textrm{in}\right\vert ^2,$
or $\vert F\vert ^2=(1-\left\vert \Gamma _l\right\vert ^2)/\left\vert 1-\Gamma _a\Gamma _{l}\right\vert ^2$
in the notation of R\&B.

We now go through the reasoning in R\&B explaining how they arrived at their result and why it differs from the treatment 
above. In this discussion we retain the subscript labels of R\&B [as in eqn.~(\ref{rbEqn8})] to facilitate direct comparison with
their work.
For the special case $\Gamma _a=\Gamma _l=0,$ both
equations above simplify to
\begin{equation}
T_\textrm{rec}=T_\textrm{sky}+T_0.
\end{equation}
In this case $T_0$ (or $T_R$ in our notation)
represents the amplitude squared of the right-moving Kurokawa noise 
wave, which is completely absorbed by the amplifier as there is no reflection. 
The left-moving Kurokawa noise wave passes through the antenna without reflection
or attenuation out into free space never to return. 

The second and third terms represent the 
contribution from the left-moving Kurokawa noise
wave, which when $\Gamma _a\ne 0$ will partially reflect off 
the antenna and contribute to
the power absorbed by the amplifier input. $T_u$ represents the component
of the left-moving Kurokawa noise wave that is uncorrelated with the right-moving
Kurowave noise wave so that the two powers add with no interference term. 
[When the left- and right-moving Kurokawa noise waves are correlated (i.e., their
relative phase is not random), $T_u$ does 
not just include $T_\textrm{unc}$ but also the 
part of the completely correlated component that does not interfere with the 
right-moving component, as we shall see further below.]

Upon the first reflection of the antenna, its power is diminished by a factor
of $\vert \Gamma _a\vert ^2,$ so the analysis may proceed as if this wave were replaced
by the right-moving wave with power $\vert \Gamma _a\vert ^2T_u.$ However when $\Gamma _l\ne 1,$
two additional factors must be added: (1) The amplitude is adjusted by a factor of $1/(1-\Gamma _a\Gamma _l)$
to account for the sum over multiple reflections back and forth between the amplifier input and the 
antenna.\footnote{This factor results from summing an infinite geometric series.}
Thus the power is adjusted by the square of the modulus of this quantity. (2)  
An additional factor of $(1-\vert \Gamma _l\vert ^2)$ to account for the fraction of power absorbed compared to
the incident right moving wave. The two factors combine to give the correction factor
\begin{equation}
\vert F\vert ^2=
\frac{
1-\vert \Gamma _l\vert ^2
}{
\vert 
1-
\Gamma _a
\Gamma _l
\vert ^2
}. 
\end{equation}
This correction factor applies equally well to the right-moving Kurokawa noise wave, 
hence the addition of the factor 
$\vert F\vert ^2$ to the $T_0$ term, which was neglected by Rogers and Bowmann. 

For the component of the left-moving Kurokawa noise wave that is completely
correlated with the right-moving Kurokawa noise wave, let
$A_L= (A_R/\vert A_R\vert)(A_{Lc}-jA_{Lr})$ be the amplitude of the left-moving 
Kurokawa wave noise source
and let 
$\Gamma _a=\vert \Gamma _a\vert (\cos \phi +j\sin \phi ).$

Then adding the amplitudes of the right-wave and once reflected 
left-wave we obtain an amplitude $A_R +\Gamma _aA_L,$
corresponding to a power 
\begin{equation}
\begin{aligned}
&\left\vert 
A_R+A_L\Gamma _a
\right\vert ^2 \\
&\quad =
\left\vert 
A_R 
\right\vert ^2
+
\left\vert 
\Gamma _a
\right\vert ^2
\left\vert A_L \right\vert ^2
%\\
%&\qquad 
+
(A_R^* A_L^{\phantom{*}})\Gamma _a
+
(A_L^*A_R^{\phantom{*}})\Gamma _a^*
\end{aligned}
\label{eqnCorr}
\end{equation}
which in turn must be multiplied by $\vert F\vert ^2.$
The first term on the second line
of eqn.~(\ref{eqnCorr}) is simply the fourth term 
$T_0$ in eqn.~(\ref{RBcorrected}),
which has already been included. The second term on the second line
can be absorbed into the second term in eqn.~(\ref{RBcorrected}).
The expression on the last line can be rewritten
\begin{equation}
\begin{aligned}
&(A_R^* A_L^{\phantom{*}})\Gamma _a
+
(A_L^*A_R^{\phantom{*}})\Gamma _a^*
\\
&\quad 
=
2\vert A_R\vert
\vert \Gamma _a\vert
\left( A_{Lc}\cos \phi + A_{Ls}\sin \phi \right)
\\
&\quad =
\vert \Gamma _a\vert
(T_c\cos \phi +T_s\sin \phi )
\end{aligned}
\end{equation}
and we thus obtain the third term in eqn.~(\ref{RBcorrected}).

For the general case of mismatch at the antenna or at the amplifier 
input, the factor 
\begin{equation}
(1-\vert \Gamma _a\vert ^2)
\vert F\vert ^2 =
\frac{
(1-\vert \Gamma _a\vert ^2)
(1-\vert \Gamma _l\vert ^2)
}{\vert 1- \Gamma _a \Gamma _l\vert ^2}
\end{equation}
in the first term of  eqn.~(\ref{RBcorrected})
is simply the familiar impedance mismatch power transfer factor $M,$
%(see Appendix \ref{mFactorAppendix}), 
defined as the ratio of
delivered power to the available power. 
Because the sky source and the amplifier noise are uncorrelated, there
is no interference.

To understand why eqn.~(\ref{RBcorrected})
rather than eqn.~(\ref{rbEqn8}) 
is correct, it is important to keep in mind that
the Kurokawa wave noise sources correspond to 
jump conditions in the amplitudes of the 
left- and right-moving waves rather than the amplitudes themselves of
the left- and right-moving waves. When this distinction is kept in
mind, there is no confusion as to how to account for the reflections
back and forth between the antenna and the amplifier. 

It has been suggested that even though the derivation leading to
eqn.~(\ref{rbEqn8}) is not correct, there may be a transformation
that could relate the two equations rendering the differences harmless 
when one uses eqn.~(\ref{rbEqn8}) rather than 
eqn.~(\ref{RBcorrected}).\footnote{Private communication}
Unfortunately this cannot be the case since the corrections
highlighted by the square brackets do depend on $\Gamma _a.$ It is crucial to correctly 
account for the differing impedance mismatch between the amplifier
on the one hand and the antenna and calibration sources on the other
hand. 

Eqns.~(\ref{rbEqn8}) and (\ref{RBcorrected}) can be reconciled if following Rogers and Bowman and progeny, one 
adopts a noise wave description where the receiver noise wave parameters depend
on $\Gamma _\textrm{in},$ which in turn depends on $\Gamma _{L}$
[see eqn.~(\ref{eqn:gammaIn})]---in other words, if
a load with a different impedance is attached to the amplifier output, the 
noise parameters change, unlike in the derivation above where the noise parameters
are a property of the amplifier 2-port and do not change when different loads
are connected to the output port. 

Rogers and Bowman imagine that the right-moving travelling wave noise source
is situated inside the amplifier
just to the right of the input port with 
an amplitude
$B_R,$ and a left-moving wave amplitude $B_L$
situated just outside the
input port. 
This runs counter to the spirit of network circuit analysis where 
a 2-port is a black box but of course is permissible as long as 
the rules of the game are clearly defined. 

The two sets of variables are related by the transformation
\begin{equation}
\begin{aligned}
B_L&=A_L+
\Gamma _\textrm{in}
A_R,\\
B_R&=
\sqrt{1-\vert \Gamma _\textrm{in}\vert ^2}
A_R
\end{aligned}
\label{eqn:RBtransform}
\end{equation}
where $A_L$ and $A_R$ are both just outside the amplifier.
Note the dependence of this transformation on $\Gamma _\textrm{in}$ (or equivalently, on $\Gamma _{l}).$ We can then use these transformations to express the Rogers \& Bowman noise wave parameters in terms of this work's noise wave parameters,
\begin{equation}
\begin{aligned}
    T_0 &= |B_R|^2 = \left| ~A_R \sqrt{1 - |\Gamma_\mathrm{in}~|^2}~\right|^2\\
    &= T_R(1 - |\Gamma_\mathrm{in}|^2),\\
    T_u &= |B_L|^2 = |A_L + \Gamma_\mathrm{in}A_R|^2\\
    &= T_L + T_R |\Gamma_\mathrm{in}|^2 + T_\mathrm{cos}\mathrm{Re}
    (\Gamma_\mathrm{in}) - T_\mathrm{sin}\mathrm{Im}(\Gamma_\mathrm{in}),\\ 
    T_c &= 2\mathrm{Re}(\langle B_R^*B_L^{\phantom{*}}\rangle) 
          =2 \mathrm{Re}
 \left\{A_R^*\sqrt{1-|\Gamma_\mathrm{in}|^2} (A_L+\Gamma_\mathrm{in}A_R)\right\}\\
   &= 2T_\mathrm{cos}\sqrt{1 - |\Gamma_\mathrm{in}|^2} 
   + 2T_R\mathrm{Re}(\Gamma_\mathrm{in})\sqrt{1 - |\Gamma_\mathrm{in}|^2},\\
    T_s &=\! -2\mathrm{Im}(\langle B_R^*B_L^{\phantom{*}}\rangle) 
          = \!-2\mathrm{Im}\left\{A_R^*\sqrt{1-|\Gamma_\mathrm{in}|^2} 
          (A_L+\Gamma_\mathrm{in}A_R)\right\}\\
    &= 2T_\mathrm{sin}\sqrt{1 - |\Gamma_\mathrm{in}|^2} - 2T_R\mathrm{Im}
    (\Gamma_\mathrm{in})\sqrt{1 - |\Gamma_\mathrm{in}|^2}.
\end{aligned}
\label{eqn:transformNVectors}
\end{equation}
The expression from Rogers and Bowman [i.e., eqn. (\ref{rbEqn8}) above] has been copied in a number
of places in the more recent global 21 cm literature, for example in 
\cite{monsalve} of the EDGES collaboration [their eqn.~(2)] and in \cite{roqueB} of
the REACH collaboration [see eqn.~(6) of their paper]. 

For experiments where the noise parameters are calibrated in situ
[e.g., REACH \citep{roqueB} and EDGES-3 \citep{cappallo}]---that is,
where $\Gamma _L$ (and thus also $\Gamma _\textrm{in}$) is fixed by the load
in place, the transformation given in eqn.~(\ref{eqn:transformNVectors}) 
relating the vectors 
$(T_0, T_u, T_c, T_s)$
and 
$(T_L, T_R, T_\textrm{cos}, T_\textrm{sin})$
has coefficients that do not depend on $\Gamma _S,$
so that two approaches are just reparameterizations 
of each other. However for experiments with
lab measured calibrations,
%(e.g., EDGES \citep{monsalve}), 
the two approaches  do not necessarily lead to the same results. 

%{\color{red}For experiments which use in-situ calibration [such as REACH \citep{roqueB} and EDGES-3 \citep{cappallo}] we would expect the Rogers and Bowman result to be valid, as opposed to lab-measured calibration such as that used by previous iterations of EDGES \citep{monsalve} where the time delay between calibration and antenna measurement could allow for considerable variations in $\Gamma_\textrm{in}$.}

\noindent
\textbf{Acknowledgements:} 
MB acknowledges support from the 
Observatoire de Paris
API SKA initiative. 
The authors thank the Kavli Foundation and the Science and Technology
Facilities Council grant for supporting the REACH project.
EdLA acknowledges the support from UKRI STFC via an Ernest Rutherford Fellowship (ST/V004425/1)
and UKRI ESPRC via a Horizon Europe Guarantee award (EP/Y02916X/1).
DdV and SP received support from the South African Radio Astronomy Observatory, which is a facility of the
National Research Foundation, an agency of the Department of Science and Technology (grant number 75322).

%BBBBB



%\bibliographystyle{mnras}

\appendix
\section{Alternative Parameterizations}

This Appendix gives the explicit relations
between the various parameterizations 
of the noise properties of an amplifier, or more abstractly
of a two-port device. 

As a consequence of eqn.~(\ref{eqn:powerWaveDef}),
the Kurokawa noise wave amplitude correlators in terms of the 
voltage and current source correlators are 
given by {\small 
\begin{equation}
\begin{aligned}
\left\langle {a_R}^* {a_R} \right\rangle &=\frac{1}{4Z_\textrm{ref}}
\Bigl[
\left\langle v^*v \right\rangle
+Z_\textrm{ref}\Bigl(
\left\langle i^*v \right\rangle
+
\left\langle v^*i \right\rangle
\Bigr)
+{Z_\textrm{ref}}^2
\left\langle i^*i \right\rangle
\Bigr] ,\\
\left\langle {a_L}^* {a_L} \right\rangle &=\frac{1}{4Z_\textrm{ref}}
\Bigl[
\left\langle v^*v \right\rangle
-Z_\textrm{ref}\Bigl(
\left\langle i^*v \right\rangle
+
\left\langle v^*i \right\rangle
\Bigr)
+{Z_\textrm{ref}}^2
\left\langle i^*i \right\rangle
\Bigr] ,\\
\left\langle {a_R}^* {a_L} \right\rangle &=\frac{1}{4Z_\textrm{ref}}
\Bigl[
\left\langle v^*v \right\rangle
+Z_\textrm{ref}\Bigl(
\left\langle i^*v \right\rangle
-
\left\langle v^*i \right\rangle
\Bigr)
-{Z_\textrm{ref}}^2
\left\langle i^*i \right\rangle
\Bigr] .
\end{aligned}
\end{equation}
}
Similarly the inverse transformation is given by
{\small \begin{equation}
\begin{aligned}
\left\langle v^*v \right\rangle &={Z_\textrm{ref}}^{\phantom{-1}}
\Bigl[
 \left\langle {a_R}^* {a_R} \right\rangle
+\left\langle {a_L}^* {a_L} \right\rangle
+\left\langle {a_L}^* {a_R} \right\rangle
+\left\langle {a_R}^* {a_L} \right\rangle
\Bigr],\\
\left\langle i^*i \right\rangle &={Z_\textrm{ref}}^{-1} 
\Bigl[
 \left\langle {a_R}^* {a_R} \right\rangle
+\left\langle {a_L}^* {a_L} \right\rangle
-\left\langle {a_L}^* {a_R} \right\rangle
-\left\langle {a_R}^* {a_L} \right\rangle
\Bigr],\\
\left\langle v^*i \right\rangle &=
\phantom{{Z_\textrm{ref}}^{-1}}
\Bigl[
 \left\langle {a_R}^* {a_R} \right\rangle
-\left\langle {a_L}^* {a_L} \right\rangle
+\left\langle {a_L}^* {a_R} \right\rangle
-\left\langle {a_R}^* {a_L} \right\rangle
\Bigr] .
\end{aligned}
\end{equation}
}

Another parameterization looks at the `noise factor,' defined as
\begin{equation}
F=\frac{
P_\textrm{n,\,source}
+
P_\textrm{n,\,amp}
}{
P_\textrm{n,\,source}
}
\end{equation}
where one assumes $P_\textrm{n,\,source}=k_BT_a(\Delta B),$
$T_a=290~K$ being the conventional ambient temperature. 
[The `noise figure' 
$\mathcal{F}=10\log _{10}(F)$
translates to a logarithmic scale.]
Here $P_\textrm{n,\,source}$
is the inherent noise of the sensor, which is assumed to arise only from Johnson noise. 
Expressing  in terms of power has the advantage that $P_\textrm{n,\,source}$ does not change when
a noiseless, lossless impedance matching 2-port is placed between between the sensor source and
the amplifier. 

We now calculate the noise power as a function of source admittance $Y,$
given by the expression [see also \cite{collin}, Sect. 10.9]
\begin{equation}
\begin{aligned}
P_\textrm{noise,\,amp}(Y)&=
\frac{
\langle v^*v\rangle YY^*
+
\langle i^*v\rangle  Y
+
\langle v^*i\rangle  Y^*
+
\langle i^*i\rangle
}{
2(Y+Y^*)
}\\
&=P_\textrm{min}+  
\frac{R_n 
\left|
Y-Y_\textrm{opt}
\right| ^2
}{
2(Y+Y^*)
}
\end{aligned}
\end{equation}
Let
\begin{equation}
z=\frac{
\langle v^*i\rangle
}{
\langle v^*v\rangle ^{1/2}
\langle i^*i\rangle ^{1/2}
}
=a+jb.
\end{equation}
It follows that 
\begin{equation}
Y_\textrm{opt}=G_n(\sqrt{1-b^2}-jb)
\end{equation}
where $G_n=
\langle i^*i\rangle ^{1/2}/\langle v^*v\rangle ^{1/2}={R_n}^{-1}$
and 
\begin{equation}
P_\textrm{min}=
\langle v^*v\rangle ^{1/2}
\langle i^*i\rangle ^{1/2}~
\frac{
\left[
\sqrt{1-b^2}+a
\right] }{
2
}.
\end{equation}
Similarly, in terms of the source impedance $Z$ we 
may obtain 
\begin{equation}
\begin{aligned}
P_\textrm{noise,\,amp}(Z)&=
P_\textrm{min}+  
\frac{G_n 
\left|
Z-Z_\textrm{opt}
\right| ^2
}{
2(Z+Z^*)
}
\end{aligned}
\end{equation}
where $Z_\textrm{opt}= {Y_\textrm{opt}}^{-1},$
thus avoiding any singularity.

\end{document}